\documentclass[onecolumn]{elsart}    % Enable this line and disable              % preceding line to obtain a two-column 
 \usepackage{natbib}
\usepackage{xcolor}
\usepackage{graphicx}          % Include this line if your 
\usepackage{amsmath}                               % you prefer.
\usepackage{amssymb}
\usepackage{svg}

\begin{document}

\begin{frontmatter}
%\runtitle{Insert a suggested running title}  % Running title for regular 
                                              % papers but only if the title  
                                              % is over 5 words. Running title 
                                              % is not shown in output.

\title{Stability Analysis of Piecewise Affine Systems with Multi-model Model Predictive Control} % Title, preferably not more 
                                                % than 10 words.
%\thanks{This work was not supported by any organization}% <-this % stops a space

\author[Manchester1]{Panagiotis Petsagkourakis}\ead{panagiotis.petsagkourakis@manchester.ac.uk},    % Add the 
\author[Manchester2]{William P. Heath}\ead{William.Heath@manchester.ac.uk},               % e-mail address 
\author[Manchester1]{Constantinos Theodoropoulos}\ead{k.theodoropoulos@manchester.ac.uk}  % (ead) as shown

\address[Manchester1]{School of Chemical Engineering and Analytical Science,The University of Manchester, M13 9PL, UK}  % Please supply                                              
\address[Manchester2]{School of Electrical and Electronic Engineering,
University of Manchester, Manchester M13 9PL, UK}        % here.

\begin{keyword}                           % Five to ten keywords,  
Unstructured uncertainty, piecewise affine, model predictive control, robust stability,                
\end{keyword}                             % keyword list or with the 
                                          % help of the Automatica 
                                          % keyword wizard

\begin{abstract} 
Constrained model predictive control (MPC) is a widely used control strategy, which employs moving horizon-based on-line optimisation to compute the optimum path of the manipulated variables. Nonlinear MPC can utilize detailed models but it is computationally expensive; on the other hand linear MPC may not be adequate. Piecewise affine
(PWA) models can describe the underlying nonlinear dynamics more accurately, therefore they can provide a viable trade-off through their use in multi-model linear MPC
configurations, which avoid integer programming. However, such schemes may introduce uncertainty affecting the closed loop stability. In this work, we propose an input to output stability analysis for closed loop systems, consisting of PWA models, where an observer and multi-model linear MPC are applied together, under unstructured uncertainty. Integral quadratic constraints (IQCs) are employed to assess the robustness of MPC under uncertainty. We create a model pool, by performing linearisation on selected transient points. All the possible uncertainties and nonlinearities (including the controller) can be introduced in the framework, assuming that they admit the appropriate IQCs, whilst the dissipation inequality can provide necessary conditions incorporating IQCs. We demonstrate the existence of static multipliers, which can reduce the conservatism of the stability analysis significantly. The proposed methodology is demonstrated through two engineering case studies.
%
%The input to output stability of the feedback interconnection of piecewise affine models with unstructured uncertainty for multi-model model predictive control is considered. In this work, the theories of dissipativity and integral quadratic constraints are combined in order to construct sufficient conditions for the input to output stability of the feedback interconnection. Specifically, the optimality conditions are initially employed in order to construct the respected integral quadratic constraints. Herein, we demonstrate the existence of static multipliers for this class of model predictive control, which can reduce the conservatism of the stability analysis significantly. The methodology is demonstrated with two engineering case studies.
\end{abstract}
 
\end{frontmatter}
%%%%%%%%%%%%%%%%%%%%%%%%%%%%%%%%%%%%%%%%%%%%%%%%%%%%%%%%%%%%%%%%%%%%%%%%%%%%%%%%
\section{Introduction}
%%%
Model predictive control (MPC) is a powerful control technique that largely relies on receding horizon-based optimization of an  objective function to compute the optimum trajectories of manipulated variables and outputs. Linear MPC has been widely used in  a number of industries~\citep{Qin2003} due to its relative simplicity and robustness~\citep{Heath2006}. Nonlinear MPC~\citep{rawlings2017model} is more appropriate for handling complex processes with underlying nonlinear dynamics. Nevertheless, nonlinear MPC can be computationally prohibitive as
computations can become slower than the process itself, making impossible to handle the process model in real time. Piecewise affine(PWA) models~\citep{BEMPORAD19991} can provide adequate accuracy for the underlying dynamical system. On the other hand, the use of PWA models in  MPC can jeopardize the computational performance as the produced optimization problem is mixed-integer programming which is NP-complete~\citep{borrelli2017predictive}. In this work a multi-model approach ~\citep{Du2015, Bonis2014} is employed to avoid mixed integer computations. Nevertheless, multi-model MPC may introduce uncertainty, which can affect the stability of the closed-loop system.

Mayne et al.~\citep{Mayne2000} has presented a complete survey of the stability and optimality conditions for MPC; however the main focus is on analysis using state terminal constraints and terminal cost for input to state stability (ISS), which can only provide local stability at the expense of additional, possibly prohibitive, complexity~\citep{Mayne2000}. Lazar et al.~\citep{Lazar2006} employed a Piecewise Quadratic (PWQ) Lyapunov function~\citep{Johansson1998} for a class of PWA MPC problems, proposing sufficient conditions for asymptotic stability with terminal constraints and cost. L{\o}vaas et al.~\citep{Løvaas2008a} have proposed a class of output robust model predictive control with all the MPC policies (within this class) satisfying a robust stability test when unstructured uncertainties are present. Alternatively, simple output feedback linear MPC with only input constraints has been verified to guarantee input to output stability~\citep{Heath2004,Heath2005,  Heath2006, Heath2010} under structured or unstructured uncertainties.  However, to best of the authors' knowledge, there is no systematic framework for analyzing the input to output stability of feedback interconnections with PWA and multi-model MPC under unstructured uncertainty. A major challenge in such a framework is to appropriately handle nonlinear and uncertain components. The theory of integral quadratic constraints (IQCs) can be used to conveniently model these components in order to construct a generic global stability analysis framework. In this work, we propose an input to output stability analysis for such feedback interconnected systems. 

Integral Quadratic Constraints first introduced by \citep{Yakubovich1967} popularized by \citep{Megretski1997}  input to output stability analysis using integral quadratic constraints (IQCs) taking advantage of input to output properties that can be adequately described by IQCs. A unified framework has been proposed in \citep{Jonsson2000} for the search of multipliers. Recently, \cite{Fetzer2017b} proposed a comprehensive analysis for the case of slope restricted nonlinearities in discrete time is presented, showing that the stability test in the literature are related.Consequently, IQCs have been widely used to perform stability and robustness analyses of dynamic systems~\citep{DAmato2001,Heath2010,Fetzer2017} in the frequency domain 
%(that can be equivalently transformed to a linear matrix inequalities (LMIs), using the KYP lemma \citep{Rantzer2012}) 
as well as in the time-domain employing  dissipativity theory ~\citep{Brogliato2007}.

%In the latter methodologies IQCs are transformed into  time domain inequalities. 
%This could be done by Parseval's theorem~\citep{DoyleR}; however infinity time horizon constraints would be produced, that are incompatible with the dissipation theory. The more generic, so-called', hard IQCs are required~\citep{Seiler2015a} that hold for any finite time horizon.

The stability theorems that have been introduced in the time-domain using dissipation theory require the existence of a positive definite matrix $P$. Pfifer and Seiler~\citep{Pfifer2015} propose a method to reduce the conservatism of the estimation of stability regions for a particular class of IQC multipliers using J-spectral factorization. Hence, it is adequate to find a symmetric matrix $P$. This approach can bring together the frequency IQC stability criterion with the dissipation approach. The J-spectral factorization can be implemented for positive-negative multipliers~\citep{Seiler2015a}. However, in the recent paper of \cite{Carrasco2017}, the equivalence between IQC and graph separation stability results is shown, if a doubly-hard factorisation is applied.

In contrast to analyses conducted in the frequency domain, time domain frameworks are not restricted to linear time invariant (LTI) systems, permitting further generalization. Robustness analysis~\citep{Pfifer2015} and robust synthesis~\citep{Wang2016} of linear parameter varying (LPV) systems using time domain IQCs have been recently developed. These results can be extended to include MPC in the dissipation inequality ~\citep{Brogliato2007}. This work focuses only on time domain IQCs allowing us to use PWA dynamics for input to output stability analysis.

\subsection{Contributions}
The main contribution of this work is to present a general framework to analyze the input-to-output stability of PWA systems for a class of multi-model MPC under unstructured uncertainty. The MPC as well as the uncertainties that arise due to the resulting model mismatches are handled by appropriate IQCs. The proposed methodology is particularly useful to analyze not only this class of PWA models but also a wider class if there exists an upper bound of the model mismatch. Four methodologies are proposed for the stability analysis (i) single parametrization (ii) conic combination, (iii) static multipliers for box constraints (iv) a combination of static multipliers with a piecewise quadratic function (PWQ) in order to reduce the conservatism even further.

\subsection{Assumptions}
For the purposes of this work we need to state the following assumptions: 
(i) Only constraints on the actuators are applied (or at least the zero solution is guaranteed feasible). 
(ii) A $\it{grey-box}$ simulator is available for the underlying process as the open loop stability is required. 
(iii) The error between the model and the real process is always bounded.

\subsection{Structure of the paper}
The paper is arranged as follows. Relevant notation is given in section 2 and the definition of the class of models addressed follows in section 3 . A brief introduction about IQCs is presented in section 4. The control scheme employed in this work is presented in section 5. All time domain IQCs for this multi-model MPC methodology are formulated in section 6 for a particular class of constraints. Section 7 introduces 4 stability theorems using the time-domain IQCs combined with the dissipation inequality for PWA models under unstructured uncertainties. Section 8, illustrates the application of the developed methodologies to 2 realistic chemical engineering case studies. Finally the conclusions and future work are given in section 9.

\section{Notations}
Let $\mathbb{Z}$ and $\mathbb{Z_{+}}$ be the set of integer numbers and positive
integer numbers including 0, respectively. $l_2^{m}$ is the Hilbert space of all square integrable and Lebesque measurable functions of size $m$, $f:\mathbb{Z_{+}} \rightarrow \mathbb{R}^{m}$. Let $l^{m}$ be the the extended space of $l_{2}^{m}$, i.e. the space of all real-valued sequences. The truncation of the function $f=f(t)$ at $T$, $f_{T}(t)$, is defined as:
\begin{equation}
f_{T}(t)=\begin{Bmatrix}
f(t) &,& \forall t\leq T\\ 
0 &,& \forall t > T
\end{Bmatrix}
\end{equation}
The function $f$ belongs to the extended space $l^{m}$ if $f_{T}(t) \in l_{2}^{m}$ for all $T>0$. $\mathbf{R}\mathbb{H}_{\infty}$ stands for the set of rational matrix transfer function matrices without poles outside the unit circle. For a complex matrix $A$, $A^{*}$ is its complex conjugate transpose. Additionally, $G^{*}$ is the $l_{2}$-adjoint operator of $G(z)\in\mathbf{R}\mathbb{H}_{\infty}$. $\left\langle f,g \right\rangle$ denotes the inner product defined as $\sum_{k=0}^{\infty}f(k)^{\intercal}g(k)=\frac{1}{\pi}\int_{-\pi}^{\pi} \widehat{f}(e^{j\omega})\widehat{g}(e^{j\omega})d\omega$. Here $\widehat{f}$ and $\widehat{g}$ denote the Fourier transforms of $f$ and $g$, respectively. The $l_{2}$ norm $\Vert f\Vert _{2}$ is defined as $\sqrt{\left\langle f,f \right\rangle}$, while $\Vert f\Vert _{1}$ is $\sum_{k=0}^{\infty}|f(k)|$. Furthermore, $G_i$ is the PWA model of $G$, $i$ being the index of each linear segment. The size of signal $x$ is indicated as $n_x$. Additionally, for the diagonal block of matrices the notation $\textnormal{diag}$ will be used, e.g. for $A$, $B$, $\textnormal{diag}(A,B)= \begin{bmatrix}A& {} \\ {} & B\end{bmatrix}$
\section{Piece-wise affine models}\label{model}
A class of piece-wise affine (PWA) models is considered in this work that given by equation~(\ref{PWA}) for every $i:\mathbb{Z_{+}}\rightarrow \mathbf{M}$, $\mathbf{M}$ being the pool of linear sub-models:
\begin{equation}\label{PWA}
\begin{split}
&x({k+1})=A_{i(k)}x({k})+B_{i(k)}u({k})+f_{i(k)}\\
&y(k) = C_{i(k)}x(k)
\end{split}
\end{equation} 
The system's sub-model changes with respect to ($y(k)$) and so it should be $i(y({k}))$ instead. However for simplicity  purposes it will be refereed as $\textit{i(k)}$. Nonlinear systems with multiple equilibrium points can accurately be  approximated by PWA models, as different linear sub-models can be utilized for different state regions. Such nonlinear systems may be found in processes described by nonlinear PDEs~\citep{Theodoropoulos2011,Bonis2012}. The available methods for constructing PWA models vary regarding each case~\citep{El-Farra2003,Bonis2014,Galan,Rewienski2003a}. The models that are exploited by the controller are as good as the collected trajectories and so the model error may destabilize the closed loop system, making the analysis in this paper crucial. Furthermore, fast model switches may destabilize~\citep{Zhang2016} the system, as a result a small perturbation or noise, which may create oscillations.

In this work, different strategies for the stability analysis of PWA models are considered, including a common storage function and a piecewise quadratic (PWQ) function \citep{Johansson1999}. These are combined with the new IQC multipliers constructed here for multi-linear MPC including uncertain/nonlinear components. The proposed methodology is found to significantly reduce  conservatism in the estimation of stability conditions. 

\section{Integral quadratic constraints}\label{secIQC}
\begin{figure}[h]
\includegraphics[scale=0.2]{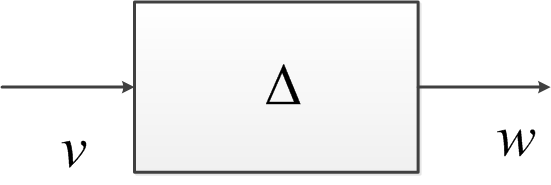}
\centering
\caption{Troublesome component analyzed by IQCs}
\label{fig:IQCs1}
\end{figure}

\begin{figure}[h!]
\centering
\includegraphics[scale=0.2]{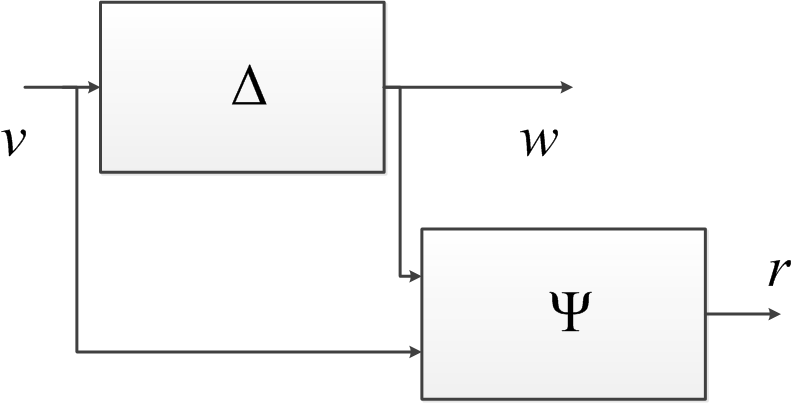}
\caption{Auxiliary system $\Psi$}
\label{fig:filters}
\end{figure}

Integral quadratic constraints (IQCs) provide a useful characterization of a given operator on a Hilbert space and they are defined in terms of self-adjoint operators with $\Pi$ being their multiplier~\citep{Megretski1997}. They are effective tools for analyzing interconnected dynamical systems consisting of uncertain, noisy or nonlinear dynamics. Integral quadratic constraints (IQCs) replace the difficult to identify or analyze components with quadratic constraints satisfied by the inputs and outputs of the troublesome component~\citep{Fetzer2017}. IQCs have been widely used for robust synthesis~\citep{Wang2016,P.Heath2016} and analysis of linear controllers~\citep{Pfifer2015}, as well as of quadratic programming controllers (MPC)~\citep{Heath2005}. The map $\Delta :l^{m}\rightarrow l^{m}$, as is shown in Fig.~\ref{fig:IQCs1}, cannot be fully specified, but some of its properties, such as monotonicity or slope-restriction, are known. Thus, $\Delta$ can be replaced by a new map $\Psi$ as in Fig.~\ref{fig:filters}~\citep{Carrasco2017}.

\begin{figure}[h!]
\centering
\includegraphics[scale=0.18]{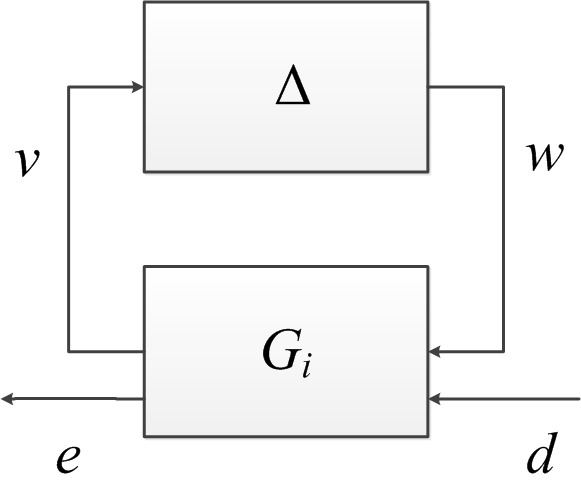}
\caption{Feedback interconnection}
\label{fig:inter}
\end{figure}

Let $\Pi$ be a bounded and self-adjoint operator; then the system's interconnection, depicted in Fig.~\ref{fig:inter}, can be described using the following equation:
\begin{align}
\begin{split}
\begin{bmatrix}
v\\e
\end{bmatrix}&=
G_{i}\begin{bmatrix}
w\\d
\end{bmatrix}\\
w&=\Delta (v)
\end{split}
\end{align}
where $G_{i}\in \mathbf{R}\mathbb{H}_{\infty}^{n_{in}\times n_{out}}$ is the transfer function matrix of the system  in equation~(\ref{PWA}), $\Delta$ is the uncertainty function, and $e$, $v$, $u$ and $w$ are the two outputs and inputs respectively. The interconnection is well posed if for each $d\in l^{m}$ and $e\in l^{m}$ there exists a unique $v\in l^{m}$ such that the map from ($d$,$e$) to ($v$,$w$) is causal~\citep{Jönsson01lecturenotes}.

%Then, it is called that $\Delta$  admits IQC defined by the multiplier $\Pi$ (or $\Delta\in {IQC}(\Pi)$)

%IQCs have been widely used for robust synthesis~\citep{Wang2016,P.Heath2016} and analysis of linear controllers~\citep{Pfifer2015}, as well as of quadratic programming controllers (MPC)~\citep{Heath2005}. 

Inequality~(\ref{IQC1}) represents a general IQC in the frequency domain. In this case it is deemed that "the uncertainty $\Delta$  admits IQC", defined by the multiplier $\Pi$ (or $\Delta\in {IQC}(\Pi)$).
\begin{equation}\label{IQC1}
	\begin{split}
&\left\langle \begin{bmatrix}
    {v}      \\
    {w} \\
\end{bmatrix},\Pi\begin{bmatrix}
    {v} \\
    {w} \\
\end{bmatrix}\right\rangle =
\int_{-\pi}^{\pi} \begin{bmatrix}
    \widehat{v}(e^{j\omega})        \\
    \widehat{w}(e^{j\omega})       \\
\end{bmatrix}^{*}\Pi(e^{j\omega})\begin{bmatrix}
    \widehat{v}(e^{j\omega})        \\
    \widehat{w}(e^{j\omega})       \\
\end{bmatrix}\geq 0
\end{split}
\end{equation}
Nevertheless, it is more convenient to use time domain analysis as nonlinear systems can be handled in a more natural way in the time domain.

\subsection{Time domain IQCs}
Time domain IQCs have been exploited for the analysis of linear parameter varying models \citep{Pfifer2015} and recently of linear time varying models~\citep{Fry2017}. Thus relevant stability criteria have been developed combining IQCs and dissipation theory~\citep{Seiler2015a}. The multiplier $\Pi$ can be factorized as $\Psi^{*} M \Psi$ and applying Parseval's theorem~\citep{DoyleR} with ${r(k)}:=\Psi~\begin{bmatrix}
{v(k)} \\ {w(k)}\end{bmatrix}$, inequality~(\ref{IQC1}) is transformed to inequality~(\ref{IQC2})
\begin{equation}\label{IQC2}
\sum_{k=0}^{\infty}r({k})^{T}Mr({k}) \geq 0
\end{equation}
Constraint (\ref{IQC2}) holds only for infinite time horizon; however the theory of dissipation requires a finite time horizon. As a result, the so-called $\it{hard}$ IQCs are necessary  (defined in~\citep{Megretski1997}) forcing the constraints to hold for every finite time horizon, $\textit{T}$ resulting in more general constraints:
\begin{equation}\label{IQC22}
\sum_{k=0}^{T}r({k})^{T}Mr({k}) \geq 0
\end{equation}
For nonlinearities varying in time, we define IQCs with a multiplier  $\textit{M}_{i}$:
\begin{equation}\label{IQC3}
\sum_{k=0}^{T}r({k})^{T}M_{i(k)}r({k}) \geq 0
\end{equation}
where $\textit{i(k)}$ corresponds to the nonlinearity $i$ at the time $k$. It should be noted that similarly to Eq.~\ref{PWA} the nonlinearity $\textit{i(k)}$ changes with respect to output ($y(k)$).

%The stability theorems that have been introduced in the time-domain using dissipation theory require the existence of a positive definite matrix $P$. Pfifer and Seiler~\citep{Pfifer2015} propose a method to reduce the conservatism of the estimation of stability regions for a particular class of IQC multipliers using J-spectral factorization. Hence, it is adequate to find a symmetric matrix $P$. This approach can bring together the frequency IQC stability criterion with the dissipation approach. The J-spectral factorization can be implemented for positive-negative multipliers~\citep{Seiler2015a}. However, in the recent paper of Carrasco and Seiler~\citep{Carrasco2017}, the equivalence between IQC and graph separation stability results is shown, if a doubly-hard factorisation is applied.

In this work, time domain hard-IQCs will be formed directly using the KKT conditions. 
\section{Model predictive control}
The quadratic programming controller exploits the multi-model scheme. The control law of the MPC consists of only input constraints (guaranteed feasibility) and it can be described  as a static nonlinearity ($\phi$) according to equation~(\ref{MPC1}) for every possible model $i$.
\begin{equation}\label{MPC1}
\begin{split}
\phi(f)=\arg \min_{U} U^{T}H_{i}U-U^{T}f&\\
s.~t.~L_{i}U\leq b&\\
M_{i}U=0&\\
b\geq 0&
\end{split}
\end{equation} 
where $U=[u_{1},\dots,u_{N_{hor}}]$ are the future input actions ($u\in \mathbb{R}^{n_{u}}$) for the control horizon $N_{hor}$, $n_{u}$ being the number of inputs. 
%\subsection{Control scheme}
%The control scheme of the problem is depicted in Fig.~\ref{fig:SCHEME}, which is equivalent to Fig.~\ref{fig:inter}, if the plant is replaced by a PWA model with model error. Therefore, $\Delta$ contains all the nonlinearities and uncertainties as $\Delta=\begin{bmatrix}
%\Delta_{unc}&{}\\{}&{\phi}
%\end{bmatrix}$ with $\Delta_{unc}$ the uncertainties or un-modelled interconnection. Hereafter, the plant will be treated as an piecewise affine model under some class of uncertainties. Hence, the uncertainties that will be included in the stability analysis later will cover a possible model error. 
%\begin{figure}[h!] 
%\includegraphics[scale=0.18]{control_scheme.png}
%\centering
%\caption{The control scheme of the current analysis}
%  \label{fig:SCHEME}
%\end{figure}       

\subsection{Output-feedback model predictive control}~\label{output}
In control schemes, all systems states are not always available, hence observers should be included in the analysis. The observer used here is Luenberger-type as used in~\citep{Heemels2008} and it can be included in the control scheme having as input the current plant's input and output. Especially, for the case of distributed parameter systems, it is almost impossible to measure all the states. Ala{\~n}a and Theodoropoulos~\citep{Alana2012} and Garcia et al.~\citep{Garcia2008} have proposed methods for finding the optimal sampling scheme and for designing observers for nonlinear distributed parameter systems. However, this work does not focus on designing an optimal observers but rather on analyzing a general closed-loop scheme.

\begin{figure}
\centering
\includegraphics[scale=0.15]{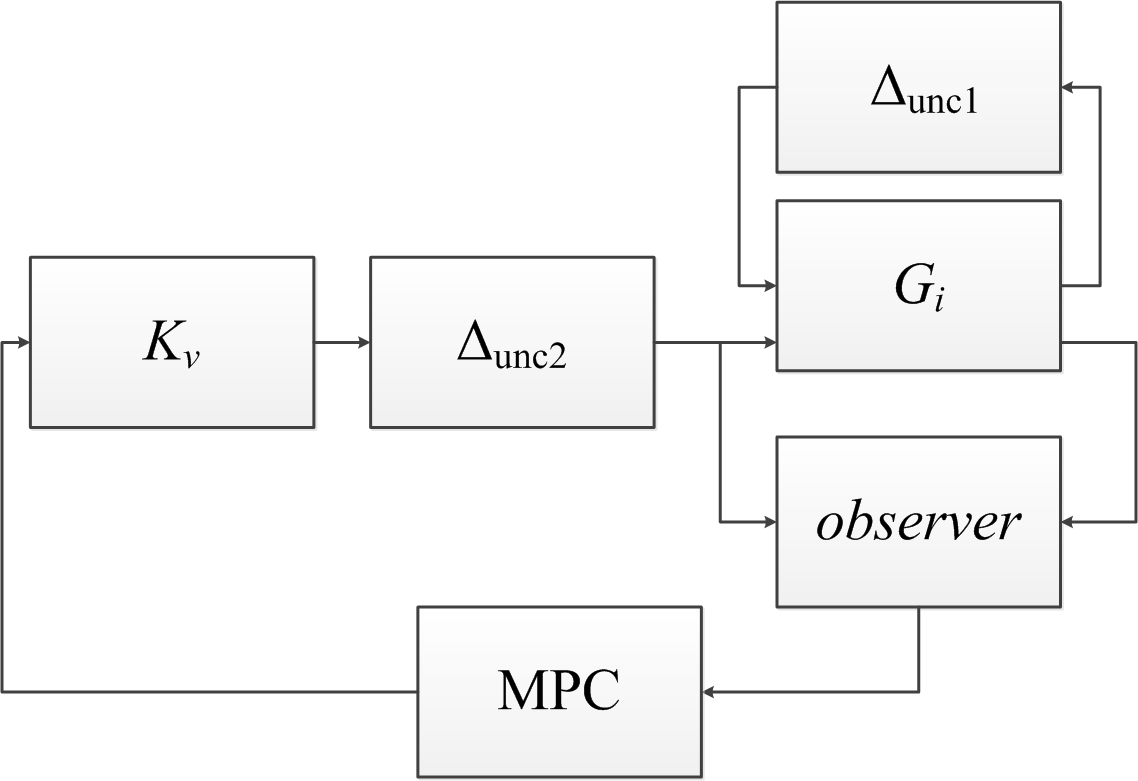}
\caption{Interconnection with the observer}
\label{fig:obs}
\end{figure}

The interconnection between the model, observer and controller is schematically depicted in Fig.~\ref{fig:obs}.
Here MPC corresponds to the nonlinearity $\phi$, while $\Delta_{unc1}$ and $\Delta_{unc2}$ represent system uncertainties such as e.g. model error. Hence, Fig.~\ref{fig:obs} can equivalently be transformed to Fig.~\ref{fig:inter}, with $\Delta = diag(Delta_{unc1},\Delta_{unc2},\phi)$ 
%
%\begin{bmatrix}\Delta_{unc1}& {} &{}\\
%{}&\Delta_{unc2}&{}\\{}&{}&\phi\end{bmatrix}$.
%
\section{Multipliers for model predictive control}

MPC Fig.~\ref{fig:obs} corresponds to nonlinear MPC for the case of a nonlinear process described by equation~(\ref{MPC1}). To perform stability analysis using IQCs, we propose to express it as a multi-linear MPC. In this case IQCs will be defined as in  Eq.\ref{IQC3} through the use of multipliers, which need to be calculated. The resulting controller can be complicated but with properties suitable for our analysis. This section will present 3 different types of multipliers for the IQCs of multi-model-based MPC. 
\subsection{Sector bounded MPC}
 For linear MPC, the input/output map is PWA~\citep{ZAFIRIOU1990}. It then follows immediately from the KKT conditions, Eq.~(\ref{MPC1}), that, provided $b\geq 0$,
\begin{equation}\label{MPC}
\phi^{\intercal}H\phi-\phi^{\intercal}f\leq 0
\end{equation}
where $\phi$ and $\textit{f}$ are the input and output of the MPC respectively \citep{Heath2004}. For constant $H$, we can integrate inequality~(\ref{MPC}), to show that the "nonlinearity" (i.e. the PWA input/output map) admits IQC:
\begin{align}
\begin{split}
&\left\langle \begin{bmatrix}
    {f}      \\
    {\phi(f)} \\
\end{bmatrix},
\begin{bmatrix}
    {O} & {I} \\
    {I} & {-2H} \\
\end{bmatrix}
\begin{bmatrix}
    {f} \\
    {\phi(f)} \\
\end{bmatrix}\right\rangle =\\
&=\langle \phi, H\phi -f\rangle \geq 0
\end{split}
\end{align}
Then, $\phi \in IQC(\Pi_{c})$ with $\Pi_{c}=\begin{bmatrix}
    {O} & {I} \\
    {I} & {-2H} \\
\end{bmatrix}$. 

This result shows that the nonlinearity is sector-bounded in the sense of~\citep{willems1971analysis}. Nevertheless,this approach is suitable only for the linear case. For our class of PWA models the $H$ in equation~(\ref{MPC1}) is now $H_{i}$. A different sub-model can be used at every sampling time with only input constraints. The controller, therefore exhibits a static nonlinearity ($\phi$) described by equation~(\ref{MPC1}) for every possible model $i$. Therefore, a new type of IQC multipliers can be introduced for this class of controllers:
\begin{lem}\label{lemma1}\citep{Petsagkourakis2017}
For every $f\in l^{n_{f}}$ and $\phi$ given by equation~(\ref{MPC1}) the nonlinearity admits the following time-domain hard IQC: 
\begin{align}\label{IQCMPC2}
\begin{split}
&\sum_{k=0}^{T} \begin{bmatrix}
    {f(k)}      \\
    {\phi(f)(k)} \\
\end{bmatrix}^\intercal
\begin{bmatrix}
    {O} & {I} \\
    {I} & {-2H_{i(k)}} \\
\end{bmatrix}
\begin{bmatrix}
    {f(k)} \\
    {\phi(f)(k)} \\
\end{bmatrix}\geq 0
\end{split}
\end{align}
\end{lem}

\begin{pf}
See~\citep{Petsagkourakis2017}.
\qed
\end{pf}
Then, the IQC from inequality~(\ref{IQCMPC2}) with multipliers 
$\begin{bmatrix}
    {O} & {I} \\
    {I} & {-2H_{i(k)}} \\
\end{bmatrix}$
is equivalent to the IQC from inequality~(\ref{IQC3}) with $\Psi=\begin{bmatrix}
{I}&{0}\\
{0}&{I}
\end{bmatrix}$. Here, the KKT conditions guarantee that inequality~(\ref{IQCMPC2}) holds for any time $T$. These will be referred to as {\it single parametrization} multipliers. %Now, this can be employed in order to provide the LMI stability conditions (see $\textit{section}$~\ref{Dis}). 

The result from Lemma~\ref{lemma1} will give conservative stability results. However, the the optimality conditions (inequality~(\ref{MPC})) can be utilized to reduce the conservatism:
\begin{lem}\label{lemma2}
For every $f\in l^{n{_{f}}}$ and $\phi$ given by equation~(\ref{MPC1}) the nonlinearity admits the following time-domain hard IQC for $\lambda_i \geq 0$:
\begin{align}\label{IQCMPC2_conic}
\begin{split}
&\sum_{k=0}^{T} \lambda_{i(k)} \begin{bmatrix}
    {f(k)}      \\
    {\phi(f)(k)} \\
\end{bmatrix}^\intercal
\begin{bmatrix}
    {O} & {I} \\
    {I} & {-2H_{i(k)}} \\
\end{bmatrix}
\begin{bmatrix}
    {f(k)} \\
    {\phi(f)(k)} \\
\end{bmatrix}\geq 0
\end{split}
\end{align}
\end{lem}
\begin{pf}
For every time interval $k$ that a model {i} is employed the following holds:
\begin{align}\label{IQCMPC2_conic}
\begin{split}
&\sum_{k=T_1}^{T_2} \begin{bmatrix}
    {f(k)}      \\
    {\phi (f)(k)} \\
\end{bmatrix}^\intercal
\begin{bmatrix}
    {O} & {I} \\
    {I} & {-2H_{i}} \\
\end{bmatrix}
\begin{bmatrix}
    {f(k)} \\
    {\phi (f)(k)} \\
\end{bmatrix}\geq 0
\end{split}
\end{align}
A conic combination can be employed such that 
\begin{align}\label{IQCMPC2_conic}
\begin{split}
&\sum_{k=0}^{T} \lambda_{i(k)} \begin{bmatrix}
    {f(k)}      \\
    {\phi(f)(k)} \\
\end{bmatrix}^\intercal
\begin{bmatrix}
    {O} & {I} \\
    {I} & {-2H_{i(k)}} \\
\end{bmatrix}
\begin{bmatrix}
    {f(k)} \\
    {\phi(f)(k)} \\
\end{bmatrix}\geq 0
\end{split}
\end{align}
%\begin{align}\label{IQCMPC2_conic}
%\begin{split}
%&\sum_{i\in M}\lambda_i\sum_{k=T_1(i)}^{T_2(i)} \begin{bmatrix}
%    {f(k)}      \\
%    {\phi(f)(k)} \\
%\end{bmatrix}^\intercal
%\begin{bmatrix}
%    {O} & {I} \\
%    {I} & {-2H_{i}} \\
%\end{bmatrix}
%\begin{bmatrix}
%    {f(k)} \\
%    {\phi(f)(k)} \\
%\end{bmatrix}\geq 0
\qed
\end{pf}
These new IQC multipliers will be termed {\it conic~combination}. We will show in section 8 that the conservatism of the stability analysis is significantly reduced through the use of conic combination multipliers.

\subsection{Multipliers for box constraints}\label{box}%%change the notation for multipliers and rewrite parts
\begin{figure}[h!]
\centering
\includegraphics[scale=0.18]{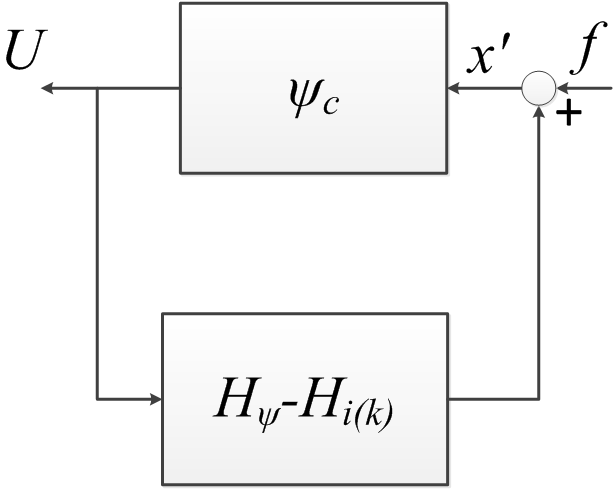}
\caption{structure of $\psi_c$}
\label{fig:psi}
\end{figure}

Here we develop a type of more general less conservative IQC multipliers for multi-model problems with a tighter class of constraints, i.e. box constraints.
A special structure of the MPC constraints is exploited by~\citep{Heath2010} for the case of linear MPC, where the existence of multipliers in the frequency domain has been demonstrated, reducing the conservatism of the analysis when fixed box constraints are utilized. We extend these results to prove the existence of static multipliers in the time domain for the multi-model case. Following the work of Heath and Li~\citep{Heath2010}, we can obtain an equivalent structure for each linear MPC corresponding to each sub-model, $i$. The resulting controller is then shown to be equivalent to a number of parallel saturation functions together with a feedback. 
Let $\psi_c$: $\mathbb{R}^{N_U}\rightarrow \mathbb{R}^{N_U}$ be the following quadratic program:
\begin{equation}\label{psi2}
\begin{split}
\psi_c(x)&=\arg \min_{U}\frac{1}{2}U^{\intercal}H_{\psi}U-U^{\intercal}x'\\
&LU\preceq b
\end{split}
\end{equation}
%\begin{lem}\label{eq1}
If we define $x'=f+(H_{\psi_c}-H_{i(k)})U$ then the feedback $U=\phi(f)$ from Eq. \ref{MPC1} is equivalent to $U=\psi_c(x')$ ~\citep{Heath2010}. The structure of $\psi_c$ is depicted in Fig.~\ref{fig:psi} for each sub-model $i$ with $N_U$ being the size of signal $U$.
%\end{lem}
%\in{pf}
%See Heath and Li~\citep{Heath2010}.\qed
%\end{pf}

The constraints $LU$ in Eq.\ref{psi2} have a specific structure for the case of box constraints. We can write $L$ and $b$ as
\begin{equation}\label{LU1}
\begin{split}
&L^{\intercal}=[L_{0}^{\intercal}, \dots , L_{N_{U-1}}^{\intercal}]\\
&b^{\intercal}=[b_{0}^{\intercal}, \dots , b_{N_{U-1}}^{\intercal}]
\end{split}
\end{equation}
with 
\begin{equation}\label{LU2}
L_{i}L_{j}^{\intercal}=0, \forall i\neq j = 0,\dots, N_{L}-1
\end{equation}
This structure is as follows: 
\begin{equation}\label{specbox}
L_i = [0,\dots,\tilde{L}_i,\dots,0]
\end{equation} 
with $\tilde{L}_i=[1, -1]^T$ and $b_i =[\bar{b} _i , -\underline{b}_i]$ with $\underline{b}_i\leq 0\leq \bar{b}_i$. Then $H_{\psi}$ can be written as 
\begin{equation}
H_{\psi_c}=\sum_{j=0}^{N_{U}-1}L_{j}^{0\intercal}\Gamma_{j}L_{j}^{0}+L^{c\intercal}\Gamma_{N_{U}}L^{c}
\end{equation}
%where $L^{0}$ and $L^{c}$ in~\citep{Heath2010} with 
where $L^{0}L^{0\intercal}=I$ and $L^{c}L^{\intercal}=0$ ~\citep{Heath2010}. The rows of $L^{0}$ form an orthonormal basis of the space spanned by the rows of $L$, and $\Gamma_{j}\in \mathbb{R}^{n_j\times n_j}$ is positive definite. 
Exploiting the orthogonality of $L_j$, we can break $\psi_c$ into several QPs. 
%%as depicted in Fig.~\ref{fig:paral}.
%%
%\begin{figure}[h!]
%%\centering
%%\includegraphics[scale=0.2]{parallelmultiplier.png}
%%\caption{structure of $\psi_c$}
%%\label{fig:paral}
%%\end{figure}
%
%To do so  $U$ is written as:
$U$ can be written as:
\begin{equation}
\label{eq2}
U=\sum_{j=0}^{N_U}u_{j}
\end{equation}
where, $j$ refers to the $j^{th}$ sub-QP of the main QP instead of the QP of each sub-model $i$. 
%
%
%\begin{lem}
%\label{eq2}
$\psi_c$ is given by (\ref{psi2}) and $U=\psi_c(x')$ from Eq.\ref{eq2}. Also \citep{Heath2010}
%as $U=\sum_{j=0}^{n_L}u_j$ with
\begin{equation}
\begin{split}
u_j&=\arg\min_{u}\dfrac{1}{2}u^{\intercal}L^{0\intercal}\Gamma_{j}L^{0}u-u^{\intercal}x\\
&L_{j}u\preceq b_{j}\\
&L_{j}^c u=0\\
&\forall j=0, \dots ,N_U-1
\end{split}
\end{equation}
Here $u$ are the degrees of freedom. Also
\begin{equation}
\begin{split}
u_{N_L}&=\arg\min_{u}\dfrac{1}{2}u^{\intercal}L^{c\intercal}\Gamma_{j}L^{c}u-u^{\intercal}x\\
&L u=0
\end{split}
\end{equation}
%
%\end{lem}%
%\begin{pf}
%See \citep{Heath2010}.\qed
%\end{pf}
%
Therefore, each $u_j$ can be written as
\begin{equation}\label{theta}
u_j(x')=L_j^{0\intercal}\theta_jL_j^0x'
\end{equation}
with $\theta_j : \mathbb{R}^{n_j}\rightarrow \mathbb{R}^{n_j}$ being the quadratic program: 
\begin{equation}\label{theta}
\begin{split}
\theta_j(p)&=\arg \min_{q} \dfrac{1}{2}q^\intercal \Gamma_j q-q^{\intercal}p\\
&L_jL_j^{0\intercal}\preceq b_j
\end{split}
\end{equation}
where $p=L_j^0 x'$ and and $q$ are the corresponding degrees of freedom. It follows immediately from the KKT conditions of Eq.~\ref{theta} that $\theta_j$ is sector bounded if $b_j\geq0$ ~\citep{Heath2007a}:
\begin{equation}
\theta_j^\intercal \Gamma_j \theta_j-\theta_j^\intercal L_j^0x'\leq 0
\end{equation}
The main result of this section can then be proven: 
%
%, where $L$ takes a specific block diagonal structure with $L=diag(\tilde{L}_0, \dots , \tilde{L}_{N_U-1})$ and $b^\intercal =[b_0^\intercal , \dots , b_{N_U-1}^\intercal]$ with $b_i =[\bar{b} _i , -\underline{b}_i]$ with $\underline{b}_i\leq 0\leq \bar{b}_i$ and $\tilde{L}_i$ has full column rank
\begin{lem}\label{lemma5}
Multipliers  $K_i=\textnormal{diag}(\kappa_{0i},\dots , \kappa_{(N_U)i})$, with $\kappa_{ji} \geq 0$  can be found for each sub-model $i\in M$, for the case of box constraints Eq.~\ref{LU1}-\ref{specbox}. If $H_\psi$ is chosen to be the identity matrix then the controller output $\phi:\mathbf{R}^{N_U}\rightarrow \mathbf{R}^{N_U}$ admits the following IQC:
\begin{align}\label{IQCMPC_new}
\begin{split}
&\sum_{k=0}^{T} \begin{bmatrix}
    {f(k)}      \\
    {\phi(f)(k)} \\
\end{bmatrix}^\intercal
M^{\phi}_i
\begin{bmatrix}
    {f(k)} \\
    {\phi(f)(k)} \\
\end{bmatrix}\geq 0
\end{split}
\end{align}
\end{lem}
%with 
\begin{equation}
M^{\phi}_i=\begin{bmatrix}
    {O} & {K_{i(k)}} \\
    {K_{i(k)}} & {-K_{i(k)} H_{i(k)}-H_{i(k)} K_{i(k)}} \\
\end{bmatrix}
\end{equation}
%%%%%pf
\begin{pf}

For each model $i\in \mathbf{M}$, $\psi_c$ admits the following time-domain IQC using a conic combination and Eq.~\ref{theta}:
\begin{align}\label{IQC_theta}
\begin{split}
&\sum_{j=0}^{N_U-1}\kappa_{ij}\sum_{k=T_1(i)}^{T_2(i)} \begin{bmatrix}
    {x'(k)}      \\
    {\psi_c(x’)(k)} \\
\end{bmatrix}^\intercal
M^{\psi_c}_j
\begin{bmatrix}
 {x'(k)}      \\
    {\psi_c(x’)(k)} \\
\end{bmatrix}\geq 0
\end{split}
\end{align}
with $M_j^{\psi_c}=\begin{bmatrix}
    {L_j^0} & {} \\
    {} & {L_j^0} \\
\end{bmatrix}^\intercal
\begin{bmatrix}
    {O} & {I} \\
    {I} & {-2\Gamma_j} \\
\end{bmatrix}
\begin{bmatrix}
    {L_j^0} & {} \\
    {} & {L_j^0} \\
\end{bmatrix}$. 

As a result, because of the orthogonality of $L^0_j$, it follows immediately for the time interval $[T_1(i),T_2(i)]$ that:
\begin{align}\label{eqsum}
\begin{split}
&\sum_{k=T_1(i)}^{T_2(i)} \begin{bmatrix}
{f(k)} \\
    {\phi(f)(k)} \\
\end{bmatrix}^\intercal
M^{\phi}_i
\begin{bmatrix}
    {f(k)} \\
    {\phi(f)(k)} \\
\end{bmatrix}\geq 0
\end{split}
\end{align}
with $$M^{\phi}_i=\begin{bmatrix}
    {-I} & {0} \\
    {H_i-H_{\psi}} & {-I} \\
\end{bmatrix}
\begin{bmatrix}
    {0} & {K_i} \\
    {K_i} & {-2K_i} \\
\end{bmatrix}
\begin{bmatrix}
    {-I} & {H_i-H_{\psi}} \\
    {0} & {-I} \\
\end{bmatrix}$$
The summation of Eq.\ref{eqsum} from 0 to $T$ gives Eq.~\ref{IQCMPC_new}
\qed
\end{pf}
The multipliers $K_i$ will be termed {\it box-constraint multipliers}. In section~\ref{app} the three types of IQC multipliers developed will be compared, to demonstrate that the box constraint multipliers produce less conservative stability results. Next, the dissipation inequality is discussed where IQC multipliers are in the stability theorems. 
\section{Dissipation inequality}
\begin{figure}%{r}{0.5\textwidth}
\centering
	\includegraphics[scale=0.13]{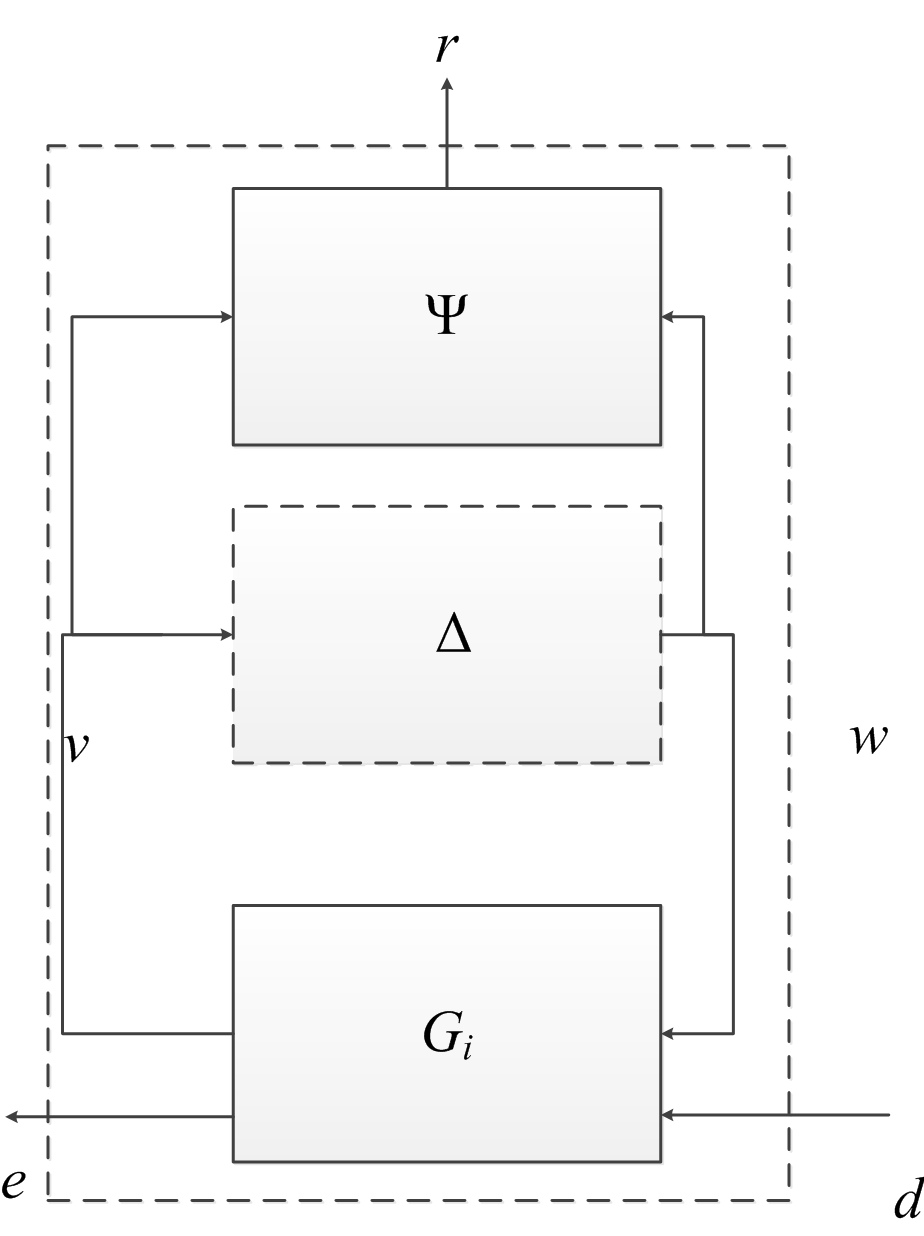}
	\caption{Extended system}
	\label{fig:ext1}
\end{figure}

The robustness of the interconnection between the dynamic system and its uncertainties(or nonlinearities) is analyzed using the extended system $\textit{G}_i^{s}$  (depicted in Fig.~\ref{fig:ext1}) where the state space vector is $x^{s}:=[ {\substack{x \\ \psi}}]$, $\psi$ being the states of $\Psi$. 
\begin{equation}\label{ext}
\begin{split}
\psi(k+1)&=A_{\psi}\psi(k)+B_{\psi_{1}}w(k)+B_{\psi_{2}}v(k)\\
z(k)&=C_{\psi}\psi(k)+D_{\psi_{1}}w(k)+D_{\psi_{2}}v(k)
\end{split}
\end{equation}
%
%where $\Psi$ is the filter of the time domain IQCs, which is equal to the identity matrix for our approach for the respected class of MPC (For other classes of IQCs see~\citep{Seiler2015a} that J-spectral factorization is applied). 
Therefore, the extended system can now be constructed:
\begin{equation}\label{ext2}
\begin{split}
x^{s}(k+1)&=A_{s}^{i}x^{s}(k)+B_{s_{1}}^{i}w(k)+B_{s_{2}}^{i}d(k)\\
z(k)&=C_{s_1}^{i}x^{s}(k)+D_{s_{11}}^{i}w(k)+D_{s_{12}}^{i}d(k)\\
e(k)&=C_{s_2}^{i}x^{s}(k)+D_{s_{21}}^{i}w(k)+D_{s_{22}}^{i}d(k)
\end{split}
\end{equation}
The structure of each matrix in Eq.~(\ref{ext2}) depends on the particular problem and on the structure of the controller. Different strategies can be implemented for MPC, e.g. two-stage integration or velocity, that affect the particular structure of the problem.
The induced controller $\textit{l}_{2}$ gain from $\textit{d}$ to $\textit{e}$ (Fig.~\ref{fig:ext1})) is defined using the interconnection between $\textit{G}_i$ and $\Delta$ in Eq.~(\ref{induced}) :
\begin{equation}\label{induced}
		||G_i,\Delta||=\sup_{d\in l_{2}} \dfrac{\Vert e\Vert}{\Vert d\Vert}	
\end{equation}
where $\Delta$ is now defined as $diag(\Delta_1,\dots,\Delta_m,\dots,\Delta_N)$ with $m\in [1,\dots,N]$ and $N$ being the number of all the uncertainties and nonlinearities in the closed loop, assumed to satisfy time domain hard IQCs (Eq.~\ref{IQC22} and~\ref{IQC3}).

%Different region can be described more accurately by different affine models, as nonlinear systems may have different behavior depending on the parameter space. Thus one affine model is not adequate to approximate the complex dynamics. However, as the controller operates, the switching between affine models may cause instabilities when the controller decides to change the model fast, thus a small perturbation may cause oscillations for an arbitrary switching rule. Therefore, the controller should be remain stable not only when there are uncertainties in the model but also for every change from model $i$ to $j$.
As mentioned above, storage functions can be used for stability analysis of PWA problems. The type of storage functions used will affect the conservatism of the stability  results \citep{Johansson1999}. A common storage function can be employed in the form of $V(x)=x^{\intercal}P x$ or a piecewise quadratic function in the form of $V(x)=x^{\intercal}P_ix$. Even though, the latter may reduce the conservatism of stability estimates, its construction usually requires significant computational time
%The number of LMIs soars
and depending on the problem we may  end up with a computationally infeasible problem. 

Here we combine a common storage function which each of the 3 IQCs developed above and provide three theorems to prove their stability analysis capabilities. 
\subsection{Stability analysis-common storage function}
\subsubsection{Single parametrization}
In subsection~\ref{output} we prove that for single-parametrization IQCs Eq.~(\ref{IQCMPC2}) holds for every $i$ and every $T$. The following theorem provides the stability boundaries of the corresponding closed loop:
\begin{thm}
Let $\textit{G}_{i} \in \mathbf{R}\mathbb{H}_{\infty}^{(n_{e}+n_{w})\times(n_{w}+n_{d})}$ be a stable system and $\Delta_m$ : $\textit{l}^{{n}_{v_m}} \rightarrow \textit{l}^{{n}_{w_m}}$ a bounded, causal operator containing every nonlinearity. The interconnection is well-posed and every $\Delta_m$ satisfies an IQC with multiplier $\Pi_m=M_m$. The controller satisfies IQC with multiplier $M_1^i$ (Lemma~\ref{lemma1}). Then $\Vert (G_{i},\Delta)\Vert<\gamma$ if there exists a symmetric matrix ${P} = P^{\intercal} \geq 0$ and nonnegative $\gamma$, $\lambda= [\lambda_1,..,\lambda_N]$ such that $LMI(\lambda,\gamma,P)$ holds. 
%\end{thm}
\begin{align} \label{LMI}
\begin{split}
&LMI(\lambda,\gamma,P):=\\
&\left[\begin{array}{ccc}
A_{s}^{i~\intercal}P A_{s}^{i}-P &  A_{s}^{i~\intercal}P B_{s1}^{i} &    A_{s}^{i~\intercal}P B_{s2}^{i}  \\
 B_{s1}^{i~\intercal}P A_{s}^{i} &  B_{s1}^{i~\intercal}P B_{s1}^{i} &    B_{s1}^{i~\intercal}P B_{s2}^{i}  \\
 B_{s2}^{i~\intercal}P A_{s}^{i} &   B_{s2}^{i~\intercal}P B_{s1}^{i} &   B_{s2}^{i~\intercal}P B_{s2}^{i}-\gamma^{2} I \end{array}
\right]+\\
&+\left[ 
\begin{array}{ccc}
C_{s2}^{i~\intercal} \\
  D_{s21}^{i~\intercal} \\
  D_{s22}^{i~\intercal}\end{array}
\right]
\left[ 
\begin{array}{ccc}
C_{s2}^{i~\intercal} \\
  D_{s21}^{i~\intercal} \\
  D_{s22}^{i~\intercal}\end{array}
\right]^{\intercal}+\\
&+\left[ 
\begin{array}{ccc}
C_{s1}^{i~\intercal} \\
  D_{s11}^{i~\intercal} \\
  D_{s12}^{i~\intercal}\end{array}
\right]
\left[ 
\begin{array}{ccc}
\lambda_1M_1^i & {\dots} & 0 \\
  {\vdots} & \ddots & {\vdots} \\
  0 & {\dots} & \lambda_N M_N\end{array}
\right]
\left[ 
\begin{array}{ccc}
C_{s1}^{i~\intercal} \\
  D_{s11}^{i~\intercal} \\
  D_{s12}^{i~\intercal}\end{array}
\right]^{\intercal}< 0
\end{split}
\end{align}
\end{thm}
\begin{pf}
Eq.\ref{LMI} can be transformed to a non-strict inequality if $\gamma^2$ is substituted by $\gamma^2 -\delta$, for a small and positive $\delta$. Multiplying then ~(\ref{LMI}) with $
\left[ x^{s\intercal}, w^{\intercal}, d^{\intercal}\right] $ and $
\left[ x^{s\intercal}, w^{\intercal}, d^{\intercal}\right] ^{\intercal}$ from the left and right respectively we get:
\begin{equation}
\begin{split}
&\lambda_{1}~r^{c}(k)^{T}M_1^{i}r^{c}(k)+\sum_{j=2}^N\lambda_{j}~r(k)^{\intercal}M_j r(k)+\\&\Delta V(k)
+e(k)^{\intercal}e(k)\leq (\gamma^{2}-\delta)d(k)^{\intercal}d(k)
\end{split}
\end{equation}
where $r^{c}$ and $r$ correspond to MPC and uncertainties respectively and storage function $\Delta V(k)=V(k+1)-V(k)$. Summing from $k=0$ to $T$ with $x^s(0)=0$ yields:
\begin{equation}
\begin{split}
&\lambda_{1}~\sum_{k=0}^{T} r^{c}(k)^{~\intercal}M_1^{i}r^{c}(k)+\sum_{j=1}^N\lambda_{j}~\sum_{k=0}^{T} r(k)^{~\intercal}M_j r(k)\\
&+V(T+1)+\sum_{k=0}^{T} e(k)^{\intercal}e(k)\leq (\gamma^{2}-\delta)~ \sum_{k=0}^{T} d(k)^{\intercal}d(k)
\end{split}
\end{equation}
The storage function is positive definite, and using the IQC conditions, inequality~(\ref{proof}) holds.
\begin{equation}\label{proof}
\sum_{k=0}^{T} e(k)^{\intercal}e(k)<\gamma^{2}~ \sum_{k=0}^{T} d(k)^{\intercal}d(k)
\end{equation}
Hence, $\Vert e\Vert< \gamma \Vert d\Vert$
\qed
\end{pf}
%%%%%%%%%%new section%%%%%%%%%%%%
\subsubsection{Conic combination}
Conservatism can be reduced by parameterizing each IQC for each affine model. The next theorem provides sufficient conditions for the closed loop stability using conic combination IQC multipliers(lemma~\ref{lemma2}). The IQCs for each MPC hold for arbitrary time, as long as we use static multipliers derived by the KKT conditions. It is worth mentioning that we can use similar arguments for every memoryless nonlinearity with static multipliers. 

\begin{thm}\label{difparam}
Let $\textit{G}_{i} \in \mathbf{R}\mathbb{H}_{\infty}^{(n_{e}+n_{w})\times(n_{w}+n_{d})}$ be a stable system and $\Delta_m$ : $\textit{l}^{{n}_{v_m}} \rightarrow \textit{l}^{{n}_{w_m}}$ a bounded, causal operator containing every nonlinearity.The interconnection is well-posed and every $\Delta_m$ satisfies IQC with multiplier $M_m$. The controller satisfies multiple IQCs given by Lemma~\ref{lemma2}. Then $\Vert (G_{i},\Delta)\Vert<\gamma$ if there exists a symmetric matrix ${P} \geq 0$ and non-negative $\gamma$, $\lambda_i = [\lambda_1^i,\lambda_2,\dots,\lambda_N]$ such that $LMI(\lambda_i,\gamma,P)$ holds.
\begin{align} \label{LMI2}
\begin{split}
&LMI(\lambda_i,\gamma,P):=\\
&\left[\begin{array}{ccc}
A_{s}^{i~\intercal}P A_{s}^{i}-P &  A_{s}^{i~\intercal}P B_{s1}^{i} &    A_{s}^{i~\intercal}P B_{s2}^{i}  \\
 B_{s1}^{i~\intercal}P A_{s}^{i} &  B_{s1}^{i~\intercal}P B_{s1}^{i} &    B_{s1}^{i~\intercal}P B_{s2}^{i}  \\
 B_{s2}^{i~\intercal}P A_{s}^{i} &   B_{s2}^{i~\intercal}P B_{s1}^{i} &   B_{s2}^{i~\intercal}P B_{s2}^{i}-\gamma^{2} I \end{array}
\right]+\\
&+\left[ 
\begin{array}{ccc}
C_{s2}^{i~\intercal} \\
  D_{s21}^{i~\intercal} \\
  D_{s22}^{i~\intercal}\end{array}
\right]
\left[ 
\begin{array}{ccc}
C_{s2}^{i~\intercal} \\
  D_{s21}^{i~\intercal} \\
  D_{s22}^{i~\intercal}\end{array}
\right]^{\intercal}+\\
&+\left[ 
\begin{array}{ccc}
C_{s1}^{i~\intercal} \\
  D_{s11}^{i~\intercal} \\
  D_{s12}^{i~\intercal}\end{array}
\right]
\left[ 
\begin{array}{ccc}
\lambda_1^iM_1^i & {\dots} & 0 \\
  {\vdots} & {\ddots} & {\vdots} \\
  0 & {\dots} & \lambda_N M_N\end{array}
\right]
\left[ 
\begin{array}{ccc}
C_{s1}^{i~\intercal} \\
  D_{s11}^{i~\intercal} \\
  D_{s12}^{i~\intercal}\end{array}
\right]^{\intercal}< 0
\end{split}
\end{align}
\end{thm}
\begin{pf}
Here conic combination is employed only for the IQC of the MPC, but it is trivial to do it for every nonlinearity. Multiplying inequality~(\ref{LMI2}) with $
\left[ x^{s\intercal}, w^{\intercal}, d^{\intercal}\right] $ and $
\left[ x^{s\intercal}, w^{\intercal}, d^{\intercal}\right] ^{\intercal}$ from the left and right respectively we get the following (for a positive $\delta$):
\begin{equation}
\begin{split}
&\lambda_{1}^i~r^{c}(k)^{~\intercal}M_1^{i}r^{c}(k)+\sum_{j=2}^N\lambda_{j}~r(k)^{\intercal}M_j r(k)+\\
&+\Delta V(k)+e(k)^{\intercal}e(k)\leq (\gamma^{2}-\delta) d(k)^{\intercal}d(k)
\end{split}
\end{equation}
Summing from $k=T_1$ to $T_2$,with $x^s(0)=0$, $[T_1 T_2]$ being the interval in which a model is employed, we have:
\begin{align}
\label{eq:ineqT1T2}
\begin{split}
&\lambda_{1}^i\sum_{k=T_1}^{T_2} r^{c}(k)^{\intercal}M_1^{i}r^{c}(k)
+\sum_{j=2}^N \lambda_{j}\sum_{k=T_1}^{T_2} r(k)^{\intercal}M r(k)^{}\\&+V({T_2+1})-V({T_1})
+\sum_{k=T_1}^{T_2} e(k)^{\intercal}e(k)\\&\leq (\gamma^{2}-\delta)~ \sum_{k=T_1}^{T_2} d(k)^{\intercal}d(k)
\end{split}
\end{align}
The summation of (\ref{eq:ineqT1T2}) over all intervals, with the storage function being positive definite and the IQC for the controller given by Lemma~\ref{lemma2}, yields:
\begin{equation}\label{proof2}
\sum_{k=0}^{T} e(k)^{\intercal}e(k)< \gamma^{2}~ \sum_{k=0}^{T} d(k)^{\intercal}d(k)
\end{equation}
from which follows that $\Vert e\Vert< \gamma \Vert d\Vert$\qed
\end{pf}
\subsubsection{Stability analysis for box constraints}\label{Dis}
For systems with box constraints the existence of IQC multipliers, $K_i$, was  proven in subsection~\ref{box}. We can then easily modify theorem~\ref{difparam} to provide stability conditions using box-constraint multipliers. 
 \begin{thm}\label{thmulti}
Let $\textit{G}_{i} \in \mathbf{R}\mathbb{H}_{\infty}^{(n_{e}+n_{w})\times(n_{w}+n_{d})}$ be a stable system and $\Delta_m$ : $\textit{l}^{{n}_{v_m}} \rightarrow \textit{l}^{{n}_{w_m}}$ a bounded, causal operator containing every nonlinearity. The interconnection is well-posed and every $\Delta_m$ satisfies IQC with multipliers $M_m$. The controller satisfies multiple IQCs with multipliers $M^{\phi}_i$ (Lemma~\ref{lemma5}). Then $\Vert (G_{i},\Delta)\Vert<\gamma$ if there exists a symmetric matrix ${P}\geq 0$ and non-negative $\gamma$, $\lambda = [\lambda_2,\dots,\lambda_N]$, and $K = diag(K_1,\dots,K_M)$ such that $LMI(\lambda,\gamma,P,K)$ holds.
\begin{equation} \label{LMImulti}
\begin{split}
&LMI(K_i, \lambda_i,\gamma,P):=\\
&\left[\begin{array}{ccc}
A_{s}^{i~\intercal}P A_{s}^{i}-P &  A_{s}^{i~\intercal}P B_{s1}^{i} &    A_{s}^{i~\intercal}P B_{s2}^{i}  \\
 B_{s1}^{i~\intercal}P A_{s}^{i} &  B_{s1}^{i~\intercal}P B_{s1}^{i} &    B_{s1}^{i~\intercal}P B_{s2}^{i}  \\
 B_{s2}^{i~\intercal}P A_{s}^{i} &   B_{s2}^{i~\intercal}P B_{s1}^{i} &   B_{s2}^{i~\intercal}P B_{s2}^{i}-\gamma^{2} I \end{array}
\right]+\\
&+\left[ 
\begin{array}{ccc}
C_{s2}^{i~\intercal} \\
  D_{s21}^{i~\intercal} \\
  D_{s22}^{i~\intercal}\end{array}
\right]
\left[ 
\begin{array}{ccc}
C_{s2}^{i~\intercal} \\
  D_{s21}^{i~\intercal} \\
  D_{s22}^{i~\intercal}\end{array}
\right]^{\intercal}+\\
&+\left[ 
\begin{array}{ccc}
C_{s1}^{i~\intercal} \\
  D_{s11}^{i~\intercal} \\
  D_{s12}^{i~\intercal}\end{array}
\right]\left[ 
\begin{array}{ccc}
M^{\phi}_i & {\dots} & 0 \\
  {\vdots} & {\ddots} & {\vdots} \\
  0 & {\dots} & \lambda_N M_N\end{array}
\right]
\left[ 
\begin{array}{ccc}
C_{s1}^{i~\intercal} \\
  D_{s11}^{i~\intercal} \\
  D_{s12}^{i~\intercal}\end{array}
\right]^{\intercal}< 0
\end{split}
\end{equation}
%
%where $M^n$ are all the nonlinearities but the MPC. 
\end{thm}
\begin{pf}
The proof is similar to the proof of theorem~\ref{thmulti} using inequality~\ref{IQCMPC_new} and it is omitted.\qed
\end{pf}
\subsection{Stability analysis using PWQ storage function}
The common storage function may yield an over-conservative approach for a PWA system. Finding a single common storage function for all the different sub-models is quite hard. This problem can be overcome by using different storage functions. Alternatively, a PWQ function can be employed. This approach may reduce the conservatism effectively, but the additional computational cost may create intractable computational problems. The following theorem provides the sufficient stability conditions when box-constraint multipliers for MPC and PWQ storage function are applied:
 \begin{thm}\label{thmulti2}
Let $\textit{G}_{i} \in \mathbf{R}\mathbb{H}_{\infty}^{(n_{e}+n_{w})\times(n_{w}+n_{d})}$ be a stable system and $\Delta_m$ : $\textit{l}^{{n}_{v_m}} \rightarrow \textit{l}^{{n}_{w_m}}$ a bounded, causal operator containing every nonlinearity. The interconnection is well-posed and every $\Delta_m$ satisfies IQC with multipliers $M_m$. The controller satisfies multiple IQCs with multipliers $M^{\phi}_i$ (Lemma~\ref{lemma5}). Then $\Vert (G_{i},\Delta)\Vert<\gamma$ if there exists a symmetric matrix ${P^j}> 0$ and non-negative  $\gamma$, $\lambda = [\lambda_2,\dots,\lambda_N]$, and $K = diag(K_1,\dots,K_M)$  such that $LMI(\lambda,\gamma,P^j,K)$ holds for all $i, j$.
%\end{thm}
\begin{align} \label{LMImultiPWQ}
\begin{split}
&LMI(K_i, \lambda_i,\gamma, P^i):=\\
&\left[\begin{array}{ccc}
A_{s}^{i~\intercal}P^j A_{s}^{i}-P^i &  A_{s}^{i~\intercal}P^j B_{s1}^{i} &    A_{s}^{i~\intercal}P^j B_{s2}^{i}  \\
 B_{s1}^{i~\intercal}P^j A_{s}^{i} &  B_{s1}^{i~\intercal}P^j B_{s1}^{i} &    B_{s1}^{i~\intercal}P^j B_{s2}^{i}  \\
 B_{s2}^{i~\intercal}P^j A_{s}^{i} &   B_{s2}^{i~\intercal}P^j B_{s1}^{i} &   B_{s2}^{i~\intercal}P^j B_{s2}^{i}-\gamma^{2} I \end{array}
\right]+\\
&+\left[ 
\begin{array}{ccc}
C_{s2}^{i~\intercal} \\
  D_{s21}^{i~\intercal} \\
  D_{s22}^{i~\intercal}\end{array}
\right]
\left[ 
\begin{array}{ccc}
C_{s2}^{i~\intercal} \\
  D_{s21}^{i~\intercal} \\
  D_{s22}^{i~\intercal}\end{array}
\right]^{\intercal}+\\
&+\left[ 
\begin{array}{ccc}
C_{s1}^{i~\intercal} \\
  D_{s11}^{i~\intercal} \\
  D_{s12}^{i~\intercal}\end{array}
\right]\left[ 
\begin{array}{ccc}
M^{\phi}_i & {\dots} & 0 \\
  {\vdots} & {\ddots} & {\vdots} \\
  0 & {\dots} & \lambda_N M_N\end{array}
\right]
\left[ 
\begin{array}{ccc}
C_{s1}^{i~\intercal} \\
  D_{s11}^{i~\intercal} \\
  D_{s12}^{i~\intercal}\end{array}
\right]^{\intercal}\leq 0
\end{split}
\end{align}
\end{thm}
\begin{pf}
The difference with theorem \ref{thmulti} is the use of the PWQ  storage function.  Hence we  only need  to prove that $\sum_{k=0}^{T}(V^{j(k)}(k+1)-V^{i(k)}(k)^)\geq 0$, $j(k)$ can be replaced by $i(k+1)$. This is easily proven as there is a fixed model for each interval. Therefore: 
\begin{equation}
\begin{split}
&\sum_{k=0}^{T}(V^{i(k+1)}(k+1)-V^{i(k)}(k))=\\
&\sum_{k=0}^{T_1}(V^{i(k+1)}(k+1)-V^{i(k)}(k))+\\
&\sum_{k=T_1+1}^{T_2}(V^{i(k+1)}(k+1)-V^{i(k)}(k))+\dots +\\
&\sum_{k=T_n+1}^{T}(V^{i(k+1)}(k+1)-V^{i(k)}(k))=\\
&V^{i(T+1)}(T+1)-V^{i(0)}(0)\geq0
\end{split}
\end{equation}
with $V^{i(0)}(0)=0$
\qed
\end{pf}
\section{Applications}\label{app}
The proposed IQC multipliers are tested for distributed parameter physical systems that are described by partial differential equations (PDEs). It is assumed that a simulator is available creating a medium size mesh and a computationally tractable finite-dimensional problem from an infinite one. In this work it is assumed that an input/output simulator-integrator is available to describe the system's dynamics accurately. Therefore, for the input $u(k)$ given the initial value $x(k)$, the next state can be computed. This form of equation is used to perform equation-free analysis for PDEs~\citep{PNAS2001} and only input to state information is available. Thus, the simulator-integrator can be described using the following equation:
\begin{equation}
x(k+1)=F(x(k),u(k))
\end{equation}
and is subsequently utilized to construct a PWA model. Here we also require that the dynamical system is open loop stable.
\subsection{Construction of the models}
It would be easy to perform successive linearizion across a sufficient number of collected trajectories.  This, however, would produce a large number of possible models and a different strategy should be employed. Here, after creating the dataset consisting of collected trajectories principle component analysis (PCA) is employed reducing the size of the  problem and avoiding issues caused by noisy data~\citep{Hastie2009},~\citep{Ding04k}. Subsequently, a clustering methodology is applied to identify data clusters and their centroids. Numerous methodologies can be applied for this procedure such as k-means~\citep{haykin2009neural}, c-means~\citep{Hastie2009} or fuzzy means~\citep{Sarimveis2002}. In this work k-means is implemented. If a computed centroid does not correspond to a feasible transient state of the physical system, the closest {\it feasible} data point is chosen. 
After reconstructing only the selected centroids, Jacobian linearizion is employed for computing the affine models. Thus only few models are created.  The procedure is described in Algorithm \ref{alg:alg} where $X$ represents the collected data of inputs and states, and $M$ is the number of clusters. 
%(Algorithms such as Fuzzy-means don't require any knowledge in advance or guess the number of corresponding clusters~\citep{sarimveis2013}).%%
%
%
This procedure typically requires 10 to 20 clusters depending on the particular problem. Furthermore, it can be combined with model reduction procedures calculating the projection basis only for the centroids. We didn't attempt to apply model reduction in this work as the focus is on the stability analysis of PWA system with unstructured uncertainties.
\begin{algorithm}[hbt!]
\caption{Computing Linear models}
  \label{alg:alg}
\textbf{Input/ Data}: $X$,$M$\\
\textbf{Output}: Model pool $\mathbf{M}$
\begin{enumerate}
\item Apply PCA to the dataset
\item Find the clusters and their centroid\\ using techniques such as k-means
\item Select the closest point to the centroid
\item Reconstruct it
\item Apply Jacobian linearizion in the full states 
\end{enumerate}
\end{algorithm} 

Algorithm 1, is used to approximate the process dynamics, while the nonlinear blocks in the closed loop are described using IQCs (see section~\ref{secIQC}). It is important to mention that any model error arising from the discertization of PDE-based model is included as an overall system error. This methodology allows to have conservative error bounds or to take into account an estimated supremum of the error. 

To illustrate the features of the proposed methodology, two illustrative case studies are considered. Firstly, the adsorption of cephalosporin in an ion-exchange resin packed-bed column ~\citep{shuler1992bioprocess} and secondly, a larger problem, a tubular reactor with an exothermic reaction~\citep{Xie2015}.

\subsection{Adsorption on an ion-exchange resin}
We apply the proposed framework to a biochemical engineering application, which is the adsorption of cephalosporin on an ion-exchange resin in a packed-bed column. The system's dynamics are described by the following differential mass conservation equations:
\begin{equation}
\epsilon\dfrac{\partial C_L}{\partial t} =-U\dfrac{\partial C_L}{\partial y}-D\dfrac{\partial^2 C_L}{\partial y^2}-(1-\epsilon)K_\alpha (C_L-C_L^{\star})
\end{equation}
where $\textit{C}_L$ is the concentration of the solute in the liquid phase, $C_L^{\star}$ is the equilibrium concentration, parameter $D=0.001 m^2/hr=7$ is the diffusion coefficient, $K_\alpha= 15 hr^{-1}$ the overall mass transfer coefficient,  $L=0.8$ the reactor length, $\epsilon=0.5$, the void fraction, and $U$ the velocity the liquid flow, which is also manipulated variable in the control problem. The equilibrium relationship together with the mass conservation equation, yield $C_L^{\star}=0.16C_L^2$ with the following boundary conditions:
\begin{align}
\begin{split}
&C_L|_{y=0} =5 
~,~\dfrac{\partial C_L}{\partial y}|_{y=L} =0
\end{split}
\end{align}
The PDE-based model was discretized in 10 finite differences using the {\it pdepe} solver in MATLAB.  
Initially, 250 trajectories were collected over a range of inlet velocities, $U$. A model pool of 14 affine sub-models was created according to Algorithm~\ref{alg:alg}.  Furthermore, it was assumed that only 5 of the 10 system states the states can be measured, hence linear observers were employed (one for each affine sub-model).
The model error was considered as a norm bounded uncertainty that admits IQC \citep{Pfifer2015}. Following the analysis of the previous sections the MPC admits IQC with all three types of multipliers described in section 6. Both common and PWQ storage functions were employed. Stability analysis was carried out, with the objective to compute the stability boundaries of the closed loop system. The objective function for the MPC problem (Eq.~(\ref{obj})), can be transformed in the form of Eq.~(\ref{MPC1}). Here, the single degree of freedom was the weight, $r$ .
%%%%%
\begin{align}\label{obj}
\begin{split}
J=\frac{1}{2}[\sum_{k=1}^{N_{out}}(y(k)-ref)^{\intercal}(y(k)-ref)+\\
r~\sum_{k=1}^{N_{in}}((u(k)-u_{ref})^{\intercal}(u(k)-u_{ref}))]
\end{split}
\end{align}
%%%%%
Here $N_{out}$=3 and $N_{in}$=2.The input variable was the velocity $U$ and box constraints were applied so $0\leq U_i\leq 5$ for each $i=1\dots N_{in}$. Therefore, the method from subsection~\ref{box} was implemented.

\begin{figure}[hbt!] 
  \centering
  \includegraphics[scale = 0.2]{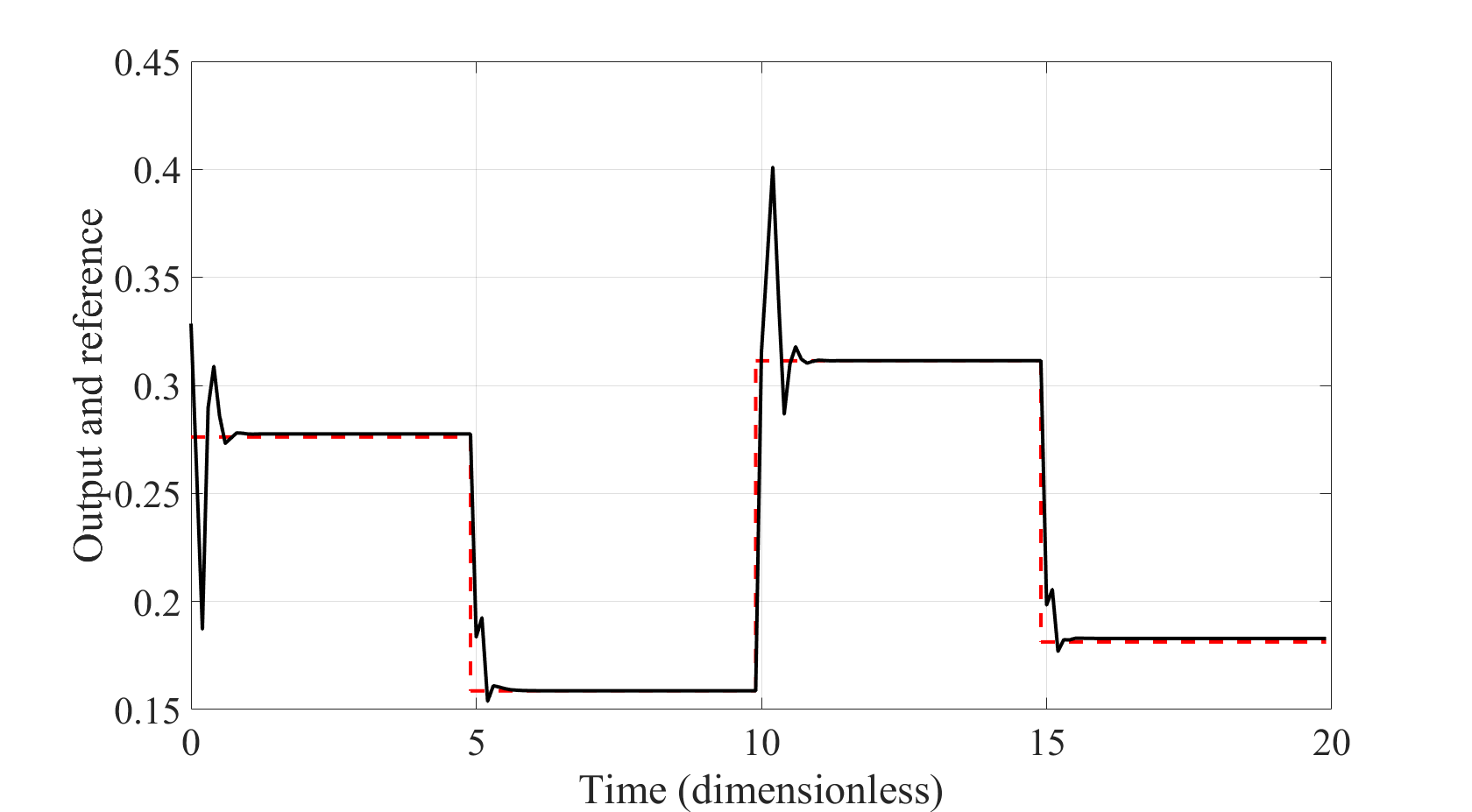}
  \caption{closed loop performance}
  \label{fig:sim1}
\end{figure}

This analysis is crucial as it can show the limits of the MPC design, since small values in the parameter $r$ can produce more aggressive controller, but may destabilize the system. In table~\ref{tab:ap1} the minimum values for $r$ are listed, for which stability can be guaranteed with the upper limit of the model's error being $b^2=0.01$ and $b^2=0.001$, respectively (Feedback uncertainty $||\Delta|| \leq b$). As it can be seen, the box-constraint multipliers reduce the conservatism of the stability boundaries, as expected. 
Additionally, it is shown that the box-constraint multipliers produce equally good results with both the common and the PWQ storage function. In addition all 3 tyoes of multipliers perform better for smaller model error, i.e. with more {\it accurate} models. 
For validation purposes we show the closed loop performance of the bio-reactor in Fig.~\ref{fig:sim1}, for $r=0.18$, where the system is confirmed to be stable. 
%The system remains stable, which can be explained by any conservatism that our method may have, but also by the fact that the model error was over-estimated to $0.01$. %Running for our framework for bound of $0.001$, the stability test reveals that the system is stable.
%
  \begin{table}[hbt!]  
  \centering
\caption{Minimum $r$ for which stability is guaranteed} \label{tab:ap1} 
\begin{tabular}{l|l|l}
	\hline
	IQC Multiplier & $\begin{matrix}
	r_{limit} \\(b^2 = 0.01)
	\end{matrix}$ & $\begin{matrix}
	r_{limit} \\(b^2 = 0.001)
	\end{matrix}$\\
	\hline
    Single Parametrization & 0.60 & 0.28 \\
	
    Conic Combination & 0.45 & 0.22\\
    
    Box-constraint & 0.28 & 0.2\\

    PWQ + Box-constraint & 0.26 & 0.18\\

	\hline
\end{tabular}
  \end{table}

\subsection{Tubular reactor}
\begin{figure}[h!]
  \centering
  \includegraphics[width=0.6\linewidth]{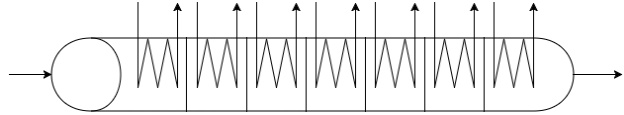}
  \caption{Tubular reactor with 8 cooling zones}
  \label{fig:tub}
\end{figure}

To further illustrate  the features of the proposed analysis, we apply the framework to a chemical engineering application, which is a tubular reactor where an irreversible exothermic reaction takes place. The system's dynamics are described with the following dimensionless equations:
\begin{align}
\begin{split}
\dfrac{\partial c}{\partial t} &=\dfrac{1}{Pe_{1}}\dfrac{\partial^{2} c}{\partial y^{2}}-\dfrac{\partial c}{\partial y}-Da~c~e^{\gamma_{1}{T}/({1+T})}\\
\dfrac{\partial T}{\partial t} &=\dfrac{1}{Pe_{2}}\dfrac{\partial^{2} T}{\partial y^{2}}-\dfrac{\partial T}{\partial y}-BDa~c~e^{\gamma_{1}{T}/({1+T})}+b(T-T_w)
\end{split}
\end{align}
where $\textit{c}$ and $\textit{T}$ are the dimensionless concentration and temperature respectively, while $\textit{T}_{w}$ is the temperature of the cooling zones, representing the degrees of freedom of the problem. In particular the cooling zones on the jacket of the reactor are separated in 8 different sectors. A schematic representation of the tubular reactor is given in Fig.~\ref{fig:tub}. The parameters of the system are $Pe_{1}=Pe_{2}$=7, $Da$=0.1, $B$=2 $b=1$, and $\gamma_{1}$=10 with the following Neumann boundary conditions:
\begin{align}
\begin{split}
\dfrac{\partial c}{\partial y}|_{y=0} &=-Pe_{1}~c~,~ 
\dfrac{\partial T}{\partial y}|_{y=0} =-Pe_{2}~T\\ 
\dfrac{\partial c}{\partial y}|_{y=L} &=0~,~
\dfrac{\partial T}{\partial y}|_{y=L} =0
\end{split}
\end{align}
The PDE-based model was discretized in 16 finite elements. The model pool is created according to Algorithm~\ref{alg:alg} as in the previous application. 180 trajectories were collected over a range of cooling temperatures, $T_w$. A model pool of 18 affine sub-models was constructed and was assumed that only 10 points out of 16 (discretization) points  along the length of the reactor can be measured.
The model error was assumed to be norm-bounded with $b^2=0.01$. Additionally, the MPC had the same design parameter, $r$, and prediction and control horizons  $N_{out}$=3 and $N_{in}$=2, respectively, as in the previous application. The input variables here (8 cooling temperatures) had upper and lower bound with $-1\leq T_{wi}\leq 1$ for $i=1\dots N_{in}$, hence the method from subsection~\ref{box} could be implemented. This case study is more computationally intensive as it has 32 states and 18 models with 8 manipulated variables for each control horizon. The inherent computational intensity of the PWQ storage function produced an intractable computational problem. Thus, only a common storage function was employed. Stability analysis was carried out, with the same objective as in the previous application.
 
Here too, as shown in Table~\ref{tab:ap2}, the box-constraint multipliers produced a substantially  less conservative stability estimate. The inclusion of PWQ storage function, on the other hand, created a computationally intractable problem. Hence, box-constraint multipliers can be used with confidence with a common storage function to produce realistic stability estimates for moderately-sized distributed parameter systems.  

\begin{figure}[h!]
  \centering
  \includegraphics[scale = 0.2]{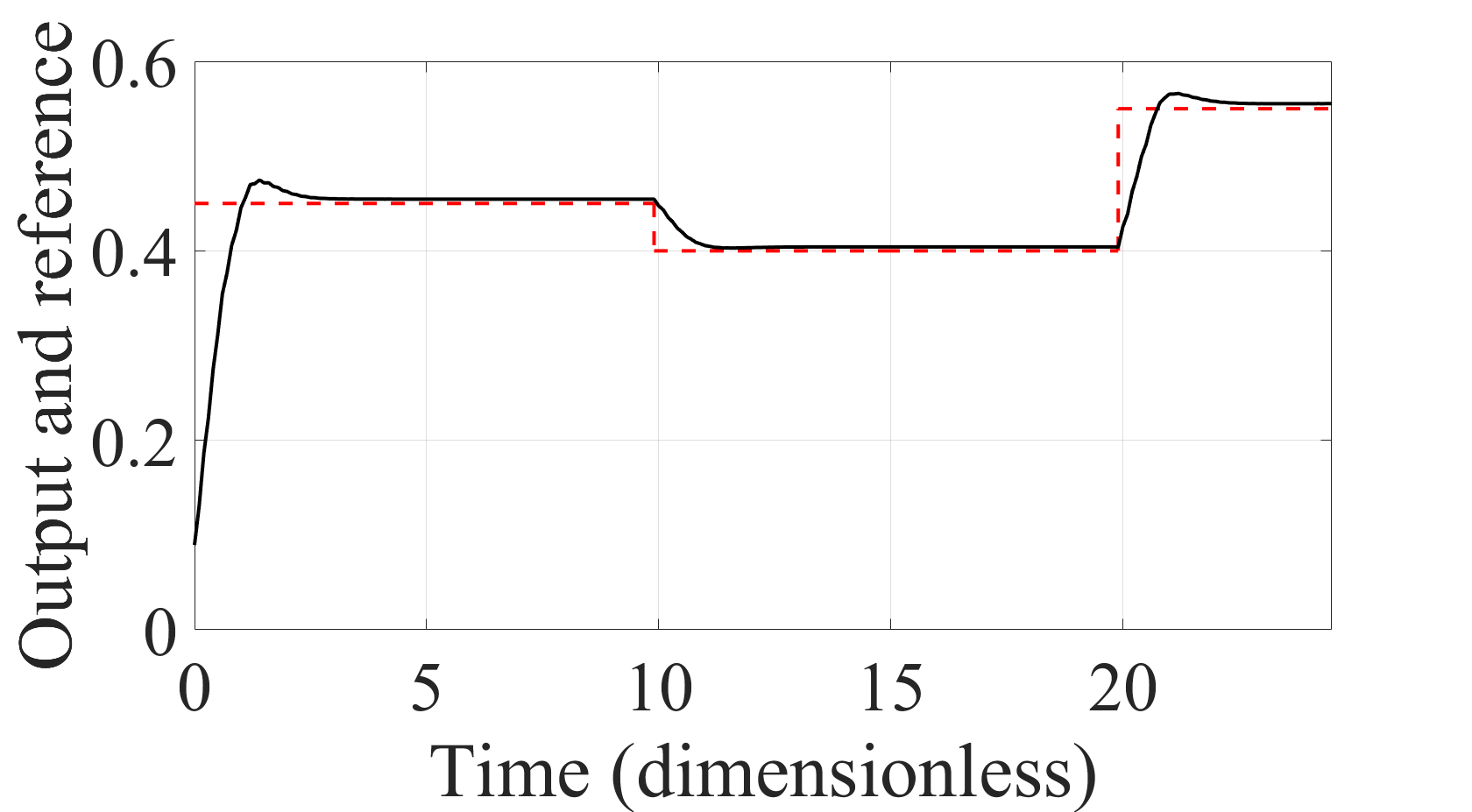}
  \caption{Closed loop performance}
  \label{fig:sim2}
\end{figure}

%Nevertheless, the method of the box constraint multipliers for each model provides a less conservative result, reducing the importance of use of PWQ storage function, that creates an intractable computational problem.  
%
  \begin{table}[h]  
  \centering
\caption{Analysis for the minimum $r$ that the stability is guaranteed} \label{tab:ap2}

\begin{tabular}{l|l}
	\hline
	IQC Multiplier & $r_{limit}$ \\
	\hline
    Single Parametrization & 0.70 \\
	
    Conic Combination & 0.25 \\
    
     Box-constraint & 0.01 \\

    PWQ + Box-constraint & - \\

	\hline
\end{tabular}

  \end{table}
For validation purposes we show the closed loop performance of the tubular reactor in Fig.~\ref{fig:sim2}, for $r=0.01$. Despite the fact that the value of $r$ is particularly small, the closed loop system is stable.

The semi-definite programming problems are all solved using MATLAB with YALMIP~\citep{Lofberg2004} and MOSEK~\citep{mosek} in computer with Intel Core i5-3570 CPU processor with 3.40GHz and 8 GB of memory. 
\section{Conclusions and Future work}
This paper focuses on the development of a robust analysis for piecewise affine models under unstructured uncertainty and multi-model-based MPC. A systematic framework was developed for accounting for uncertainties such as model error. Sufficient conditions were presented for PWA models using three different type of IQC multipliers for the controller's nonlinearity in conjunction with common and PWQ storage functions. It was shown, through two illustrative case studies, that box-constraint multipliers significantly reduce conservatism in the prediction of stability boundaries.  For the absorption column with 14 sub-models and 10 states the single parametrization multipliers with a common storage function required 11 cpu-sec per each $r$, and 5.3 cpu-min when the PWQ storage function was employed. For the tubular reactor with 18 sub-models and 32 states, around 40 cpu-min are required for the box-constraint multipliers with a common storage function. The available computer memory was not enough to solve the problem with the PWQ storage function. Therefore, the difference is substantial when the number of states increases and additional work needs to be performed in regards to the handling of large-scale systems. In a future work, model order reduction will be employed ~\citep{Theodoropoulos2011,Luna-Ortiz2005,Xie2011} to describe the infinite dimension system as finite reduced models. This is the first time that IQCs have been used beyond the scope of linear MPC and we believe this is a significant step towards their use for the analysis of complex nonlinear systems.     
\section*{Acknowledgments}
The University of Manchester Presidential Doctoral Scholarship Award to PP is gratefully acknowledged.

\bibliography{all2}           % and a bib file to produce the 
                                 % bibliography (preferred). The
                                 % correct style is generated by
                                 % Elsevier at the time of printing.

%\begin{thebibliography}{99}     % Otherwise use the  
                                 % thebibliography environment.
                                 % Insert the full references here.
                                 % See a recent issue of Automatica 
                                 % for the style.
%  \bibitem[Heritage, 1992]{Heritage:92}
%     (1992) {\it The American Heritage. 
%     Dictionary of the American Language.}
%     Houghton Mifflin Company.
%  \bibitem[Able, 1956]{Abl:56}
%     B.~C.~Able (1956). Nucleic acid content of macroscope. 
%     {\it Nature 2}, 7--9. 
%  \bibitem[Able {\em et al.}, 1954]{AbTaRu:54}   
%     B.~C. Able, R.~A. Tagg, and M.~Rush (1954).
%     Enzyme-catalyzed cellular transanimations.
%     In A.~F.~Round, editor, 
%     {\it Advances in Enzymology Vol. 2} (125--247). 
%     New York, Academic Press.
%  \bibitem[R.~Keohane, 1958]{Keo:58}
%     R.~Keohane (1958).
%     {\it Power and Interdependence: 
%     World Politics in Transition.}
%     Boston, Little, Brown \& Co.
%  \bibitem[Powers, 1985]{Pow:85}
%     T.~Powers (1985).
%     Is there a way out?
%     {\it Harpers, June 1985}, 35--47.

%\end{thebibliography}

\end{document}